\documentclass[prd,nofootinbib,tightenlines]{revtex4}
\usepackage{graphicx}
\def\journal#1, #2, #3#4, #5#6#7#8    {
    {#1~} {#2}  (#5#6#7#8) #3#4}

\usepackage{amssymb}
\usepackage{amsmath}
\usepackage{lscape}
\begin{document}


\renewcommand{\thesection}{\arabic{section}}
\renewcommand{\theequation}{\arabic{equation}}
\renewcommand {\c}  {\'{c}}
\newcommand {\cc} {\v{c}}
\newcommand {\s}  {\v{s}}
\newcommand {\CC} {\v{C}}
\newcommand {\C}  {\'{C}}
\newcommand {\Z}  {\v{Z}}
\newcommand{\pv}[1]{{-  \hspace {-4.0mm} #1}}

\newcommand{\be}{\begin{equation}} \newcommand{\ee}{\end{equation}}
\newcommand{\bea}{\begin{eqnarray}}\newcommand{\eea}{\end{eqnarray}}
\newcommand{\grad}{\bm \nabla}

\baselineskip=14pt

\begin{flushright}
\hfill{SINP/TNP/2008/07}\\
\end{flushright}

\begin{center}
{\bf  \Large Inequivalent quantization of the rational Calogero model \\
with a Coulomb type interaction} 
\bigskip

B. Basu-Mallick{\footnote{e-mail: biru@theory.saha.ernet.in}}, 
Kumar S. Gupta {\footnote{e-mail:kumars.gupta@saha.ac.in}}\\
Theory Division, Saha Institute of Nuclear Physics, 1/AF Bidhannagar, Calcutta 700064, India\\[3mm]

\bigskip

S. Meljanac {\footnote{e-mail: meljanac@irb.hr}}, A. Samsarov {\footnote{e-mail: asamsarov@irb.hr}} \\  
 Rudjer Bo\v{s}kovi\'c Institute, Bijeni\v cka  c.54, HR-10002 Zagreb,
Croatia \\[3mm]

\end{center}
\setcounter{page}{1}
\bigskip


\begin{abstract}
We consider the inequivalent quantizations of a $N$-body rational Calogero
model with a Coulomb type interaction. It is shown that for certain range of the coupling constants, this
system admits a one-parameter family of self-adjoint extensions. We analyze both the bound 
and scattering state sectors and find novel solutions of this model. We also
find the ladder operators for this system, with which the previously known solutions can be 
constructed.

\bigskip 
\noindent
PACS number(s): 02.30.Ik, 03.65.Fd, 03.65.-w \\
\bigskip
\bigskip
Keywords: calogero model, self-adjoint extension
\end{abstract}


\maketitle
 


\section{Introduction}

Exactly solvable quantum many body systems like Calogero model and its 
variants \cite{Calo,BS,pr}   have found diverse applications in 
many branches of contemporary physics, including
generalized exclusion statistics \cite{poly},
quantum hall effect \cite{qhe}, Tomonaga-Luttinger liquid \cite{ll}, quantum
chaos \cite{rmt},  quantum electric transport in mesoscopic system \cite{qet},
spin-chain models \cite{hs1,hs2}, Seiberg-Witten theory \cite{sw} and
black holes \cite{black}.
The rational Calogero model is described by $N$ identical particles interacting
with each other through a long-range inverse-square and harmonic
interaction on the line \cite{Calo}. 
The exact spectrum of this rational Calogero model with harmonic 
confinement has been found through a variety of different techniques 
\cite{Calo,Meljanac:2006uf}, all of which impose the boundary condition
that the wavefunction and the current vanish when any two or more particles 
coincide. With this boundary condition the Hamiltonian is self-adjoint,
which ensures the reality of eigenvalues as well as the completeness of the 
states.  However it was found later that, within a certain region of the 
parameter space,  there exist more general 
boundary conditions for which the rational Calogero Hamiltonian
(with and without harmonic confinement) admits 
self-adjoint extensions and yields a rich variety of spectra \cite{we, feher}. 
As is well known, the possible boundary conditions for an operator are 
encoded in the choice of its domains, which are classified by the self-adjoint
extensions \cite{reed} of the operator. Such 
self-adjoint extensions play important roles in a variety of physical contexts
including Aharonov-Bohm effect \cite{gerbert}, two and three dimensional
delta function potentials \cite{jackiw}, anyons \cite{manuel}, anomalies
\cite{esteve}, $\zeta$-function renormalization \cite{falo},
particle statistics in one dimension \cite{bal} and black
holes \cite{trg}. So it should be interesting to find out more 
examples of exactly solvable models which can be quantized by using 
the method of self-adjoint extension.

In this context it may be noted that an exactly solvable variant 
of the rational Calogero model has been constructed by Khare \cite{khare}, 
where the confining simple harmonic potential is replaced 
by a coulomb-like interaction. 
The bound states of this model can be related to those of the 
rational Calogero model with harmonic confinement by using 
the underlying $SU(1,1)$ algebra \cite{ghosh}. However, apart from 
having an infinite number of bound states, 
this model with coulomb-like interaction also supports
continuous scattering states. Similar to the case of original Calogero model,
these bound as well as scattering states have been constructed by using 
the boundary condition that wavefunction and the current vanish
when any two or more particles coincide. 
Due to a factorization property of the eigenfunctions, 
the eigenvalue problem of this many body system 
can be reduced to that of the corresponding 
radial Hamiltonian.  In this article, our aim is to find out more general 
boundary conditions which admit self-adjoint extensions for the 
radial part of the rational Calogero model with coulomb-like interaction
and study the related spectra. 
   
The arrangement of this article is as follows. In Sec.2 we briefly 
recapitulate how to separate the radial part of the rational Calogero 
Hamiltonian with coulomb-like interaction. 
Then we discuss about the most 
general form of eigenstates associated with this radial Hamiltonian $H_r$.
These eigenstates could be singular or nonsingular at $r=0$ value 
of the radial coordinate $r$. The bound and scattering states found 
by Khare are all nonsingular at $r=0$ \cite{khare}.  In Sec.3 we show that 
such nonsingular bound states can also be constructed by using 
creation annihilation operators associated with the 
underlying $SU(1,1)$ algebra. In Sec.4 we show that the 
radial Hamiltonian $H_r$ admits self-adjoint extensions within 
a certain region of the parameter space. Inequivalent quantizations 
of $H_r$ by using this method lead to bound and 
scattering eigenstates which are in general singular at $r=0$.  We explicitly 
construct such bound states and scattering states in Sec.5 and Sec.6 
respectively. In Sec.6, we  also derive the scattering matrix for 
the scattering states and show that the eigenvalues of the bound states 
can be reproduced from the poles of this scattering matrix. 
Sec.7 is the concluding section.


\section{General form of eigenfunctions of the radial Hamiltonian}

 The Hamiltonian for the rational Calogero model with the Coulomb type
 term is given by
\begin{equation} \label{hamiltonian}
  H = - \sum\limits_{i=1}^{N} \frac{{\partial}^{2}}{\partial x_{i}^{2}}
      + g \, \sum\limits_{i < j} \frac{1}{{( x_{i} - x_{j})}^{2}} -
        \frac{\alpha}{ \sqrt{ {\sum}_{i < j} {( x_{i} - x_{j})}^{2} } } \, .
\end{equation}
It describes the $\; N$-body problem with equal mass in $1$-dimension
and in units such that $\; 2m {h}^{-2} =1. \;$
It is also understood that the coupling constant
$\; g \;$ is constrained to satisfy condition
 $\; g > -\frac{1}{2}. \;$ The parameter $\; \alpha \;$
will be allowed to have any value, positive or negative, which will
provide us with the possibility to treat the attractive as well as the 
repulsive Coulomb potential, respectively.
This point is somewhat different in comparison to the treatment made in \cite{khare}
where only the case for $\; \alpha > 0 \;$ is considered.
Having the Hamiltonian (\ref{hamiltonian}), we intend to solve the
eigenvalue problem
\begin{equation} \label{fulleigenvalueeq}
  H \Psi = E \Psi.
\end{equation}
Following \cite{Calo}, we consider the above eigenvalue equation in a sector of
configuration space corresponding to a definite ordering of particles
given by $\; x_{1} \geq x_{2} \geq \cdot \cdot \cdot \geq x_{N}. \;$
The translationaly invariant eigenfunctions of the Hamiltonian $\; H \;$
can be factorized as
\begin{equation} \label{factorization}
  \Psi =  \prod\limits_{i < j} {( x_{i} - x_{j})}^{a + \frac{1}{2}} 
  \phi (r) P_{k} (x), 
\end{equation}                  
where $\; x \;$ is an abbreviation for 
$\; \{ x_{1}, x_{2}, \cdot \cdot \cdot, x_{N} \},  \;$ 
$\; a =\pm \frac{1}{2} \sqrt{1 + 2g} \; $
and $\; r \; $ is the collective radial variable defined as
\begin{equation} \label{collectiveradialvariable}
  r^{2} = \frac{1}{N} \sum\limits_{i < j} {( x_{i} - x_{j})}^{2}.
\end{equation} 
Functions $\; P_{k} (x) \; $ are translationaly invariant,
homogeneous polynomials of degree $\; k, \; k \geq 0. \;$
They satisfy the equation
\begin{equation} \label{hompoleq}
   \bigg [\sum\limits_{i=1}^{N} \frac{{\partial}^{2}}{\partial x_{i}^{2}}
      +2(a + \frac{1}{2})\sum\limits_{i < j} \frac{1}{( x_{i} - x_{j})}
     (\frac{\partial}{\partial x_{i}} - \frac{\partial}{\partial x_{j}} ) \bigg ] P_{k} (x)
      = 0.
\end{equation}
This equation is analyzed in detail by Calogero \cite{Calo}. After inserting
the factorized form of $\Psi$ from Eq. (\ref{factorization}) 
into Eq.(\ref{fulleigenvalueeq}) and
using Eqs.(\ref{collectiveradialvariable}) and (\ref{hompoleq}), we get an
equation for $\phi(r)$ as 
\begin{equation} \label{eigenvalueeq}
  H_{r} ~ \phi = E \phi,
\end{equation}
where 
\begin{equation} \label{radialhamiltonianfirst}
 H_{r} = -\frac{d^2}{d r^2}  - (2k + 2b + 1) \frac{1}{r} \frac{d}{d r}
  - \frac{\alpha}{\sqrt{N} r}. 
\end{equation}
The parameter $\; b \; $ which enters Eq.(\ref{radialhamiltonianfirst}) is defined as
\begin{equation} \label{parameterb}
  b = \frac{N(N-1)}{2} \big(a+\frac{1}{2}\big) + 
  \frac{N}{2} - \frac{3}{2}.
\end{equation}
It should be noted that, due to Eq. (\ref{factorization}),  the measure 
of the quadratically integrable $\phi (r)$ is given by
$ \; d \sigma = r^{\beta} dr, \; $ where $ \; \beta = 2k + 2b + 1. \; $
Thus, the eigenvalue equation (\ref{fulleigenvalueeq})
of the Calogero model with Coulomb-like 
interaction is reduced to the eigenvalue equation 
(\ref{eigenvalueeq}) of the corresponding
 radial Hamiltonian $H_r$, which may be explicitly written as  
\begin{equation} \label{before1}
 -\frac{d^2}{d r^2} \phi (r) - (2k + 2b + 1) \frac{1}{r} \frac{d}{d r}
 \phi (r) - \frac{\alpha}{\sqrt{N} r} \phi (r) = E \phi (r).
\end{equation}
However, for the purpose of finding out all possible  
boundary conditions on the wavefunctions for which $H_r$ admits 
self-adjoint extensions, in due course we shall also need to study the 
deficiency subspaces for this radial Hamiltonian.
Thus for the sake of convenience we consider a somewhat more
general form of Eq.(\ref{before1}), namely,
\begin{equation} \label{1}
 -\frac{d^2}{d r^2} \phi (r) - (2k + 2b + 1) \frac{1}{r} \frac{d}{d r}
 \phi (r) - \frac{\alpha}{\sqrt{N} r} \phi (r) = \tilde{E} \phi (r),
\end{equation}
with $ \; \tilde{E} = E, +i, -i, \;$ depending on whether we are
interested in the eigenfunctions or the deficiency
subspaces for the radial Hamiltonian (\ref{radialhamiltonianfirst}).

By performing  the following transformations 
\begin{equation}  \label{2}
  \phi (r) = r^{- \frac{\beta}{2}} \psi (r),
 \; \beta = 2k + 2b +1, \quad   y = cr,
\end{equation}
with the parameter $c$ as yet unspecified constant, Eq.(\ref{1}) can be reduced to Whittaker's equation, which in turn
is related to confluent hypergeometric equation of the form
\begin{equation} \label{3}
 \left [ y \frac{d^2}{d y^2} + (2\mu + 1 - y) \frac{d}{dy} 
 - \left (\frac{1}{2} + \mu - \kappa \right ) \right ] \chi (y) = 0.
\end{equation}
The last step, where Whittaker's equation, satisfied by the function
$\; \psi, \;$ reduces  to
Eq.(\ref{3}), is performed  
by means of the factorization $\; \psi(r) = e^{-\frac{y}{2}}
y^{\frac{\beta}{2}} \chi (y). \; $
The parameters introduced in Eq.(\ref{3}) are given as
$$
  \mu = \frac{\beta}{2} - \frac{1}{2},
$$
\begin{equation} \label{4}
  c = 2\sqrt{- \tilde{E}},
\end{equation}
$$
 \kappa = \frac{\alpha}{c \sqrt{N}} = \frac{\alpha}{ \sqrt{-4N \tilde{E} }}.
$$
The general solution to Eq.(\ref{3}) is known to be the linear
combination of the confluent hypergeometric functions of the first and
the second kind \cite{abr}
\begin{equation} \label{5}
  \chi (r) = A \, M  \left ( \frac{1}{2} + \mu - \kappa,~ 2\mu + 1,~ c r
  \right )
   + B \, U  \left ( \frac{1}{2} + \mu - \kappa,~ 2\mu + 1,~ c r \right ),
\end{equation}
where $A$ and $B$ are arbitrary constants. Due to the relations 
$\; \psi (r)= e^{-\frac{y}{2}} y^{\frac{\beta}{2}} \chi(y) \; $ and $y=cr$, 
Eq.(\ref{5}) yields the general form of $\psi (r)$ as 
\begin{equation} \label{6}
  \psi(r) = e^{-\frac{cr}{2}} {(cr)}^{\frac{\beta}{2}}
 \left (   A \, M \left ( \frac{\beta}{2} - \kappa,~ \beta,~ c r \right )
  + B \, U \left ( \frac{\beta}{2} - \kappa,~ \beta,~ c r \right ) \right ).
\end{equation}
Since the measure on the space of quadratically integrable $\phi $ functions
is $ d \sigma = r^{\beta} dr, $ Eq.(\ref{2}) implies that we have
$$
 \int {\phi}^{\star} \phi d \sigma = \int {\psi}^{\star} \psi dr,
$$
showing that on the space of quadratically integrable $\psi $ functions
the measure is simply $dr$.

It may be noted that, in terms of some generic parameters 
$a,b \;$and the variable $z,$ the
 confluent hypergeometric functions $M$ and $U$ are
determined by the expressions \cite{abr}
\begin{equation} \label{7}
   M (a,b,z)= 1+ \frac{az}{b} + \frac{a(a+1)z^2}{b(b+1) 2!}+
   \cdot \cdot \cdot + \frac{(a)_n z^n}{(b)_n n!}+ \cdot \cdot \cdot,
\end{equation}
\begin{equation} \label{8}
   U (a,b,z ) = \frac{\pi}{\sin \pi b}
 \bigg [ \frac{M ( a,b,z )}
{\Gamma (1+a-b) \Gamma (b)}  -   ( z )^{1-b}
\frac{M (1+a-b, 2-b, z )}
{\Gamma (a) \Gamma (2-b)} \bigg ],
\end{equation}
where the symbol $ \; (a)_n \; $ means
\begin{equation}
(a)_n = a (a+1) (a+2) ...... (a+n-1),~(a)_0 = 1.
\end{equation}
These expressions show that while
$M (a,b,z)$ is nonsingular at $z=0$,  $U (a,b,z )$ could be  
singular at $z=0$ with leading power $z^{1-b}$.
Thus, for $B\neq 0$, the solution of $\psi (r)$ given 
in Eq.(\ref{6}) is singular at $r=0$. 
In this article, our main aim is to employ such singular solutions
of $\psi (r)$ to construct bound and scattering states 
through the method of self-adjoint extension.  On the other hand 
it may be observed that, the solution of $\psi (r)$ given in Eq.(\ref{6})
would be nonsingular at $r=0$ for the case
$B=0$ and $\beta \geq 0$.  
Bound and scattering state 
solutions found in \mbox{Ref. \cite{khare}} all correspond to such nonsingular 
solutions of $\psi (r)$.


\section{Construction of nonsingular bound states through ladder operators}

Let us now concentrate on the set of bound states corresponding to the
spectrum found by Khare. These states are solutions to Eq.(\ref{1})
when $\tilde{E} = E, \; E < 0$. In this case, parameters $\; c \;$
and $\; \kappa, \; $ introduced in (\ref{4}), become real parameters
 $\; c = 2 \sqrt{|E|} \;$  and $\; \kappa = \frac{\alpha}{\sqrt{4N
  |E|}}.$ Since $U$ is singular at $r = 0,$ after
utilizing equation (\ref{2}), we are left
with the wavefunctions of the form
\begin{equation} \label{30}
 \phi (r) = r^{- \frac{\beta}{2}} e^{- \frac{1}{2} cr}
 {(cr)}^{\frac{\beta}{2}} M (\frac{\beta}{2} - \kappa,~  \beta,~  cr).
\end{equation}
Due to the fact that $M$ comprises an infinite diverging series, it
has to be truncated and this can be achieved by setting   $\;
\frac{\beta}{2} - \kappa = -n, \quad n = 0,1,2,3,...  \;$ With this
truncation condition
$M$ reduces to the associated Laguerre polynomials, 
$\; L_{n}^{(\alpha)} (x) = \frac{{(\alpha + 1)}_{n}}{n!} M(-n,~  \alpha + 1,~ 
x),  \;$ and wavefunctions (\ref{30}) belong to the following discrete set of
normalized eigenfunctions, labeled by the quantum number $n,$ 
\begin{equation} \label{31}
 {\phi}_{n} (r) = \sqrt{\frac{n!}{(2n + \beta) \Gamma (n + \beta)}}
 {c}^{\frac{\beta + 1}{2}} e^{- \frac{1}{2} cr}
 L_{n}^{(\beta - 1)} (cr) \equiv C_{n} e^{- \frac{1}{2} cr}
 L_{n}^{(\beta - 1)} (cr) ,  \quad n = 0,1,2,3,...
\end{equation}
They are normalized to unity with respect to the measure $ \; d \sigma =
r^{\beta} dr, \;$ i.e. they satisfy $\; \int {\phi}_{n}^{\star} {\phi}_{n} d \sigma =1.  \;$
The corresponding bound state energies follow from the aforementioned
truncation condition and are given as
\begin{equation} \label{32}
 E_{n} = - \frac{1}{4N} \frac{{\alpha}^2}{{(k + b + n + \frac{1}{2})}^2},  \quad n = 0,1,2,3,...
\end{equation}

By using recursive relations for the associated Laguerre functions
$$
 (n + \alpha) L_{n-1}^{(\alpha)} (x) = (2n + 1 + \alpha - x)
 L_{n}^{(\alpha)} (x)   - (n+1) L_{n+1}^{(\alpha)} (x),
$$
\begin{equation} \label{33}
  x \frac{d}{dx} L_{n}^{(\alpha)} (x) = n L_{n}^{(\alpha)} (x) - (n +
  \alpha) L_{n-1}^{(\alpha)} (x),
\end{equation}
$$
  x \frac{d}{dx} L_{n}^{(\alpha)} (x) + ( n + 1 + \alpha - x)  L_{n}^{(\alpha)} (x) =
 (n + 1) L_{n+1}^{(\alpha)} (x),
$$
we can find recursions
\begin{equation} \label{34a}
  \left ( n - \frac{1}{2} y - y \frac{d}{dy} \right ) {\phi}_{n} (y) = 
  \sqrt{\frac{n(2n + \beta -2)(n + \beta -1)}{2n + \beta}} {\phi}_{n-1} (y),
\end{equation}
\begin{equation} \label{34b}
 \left ( n + \beta - \frac{1}{2} y + y \frac{d}{dy} \right ) {\phi}_{n} (y) = 
  \sqrt{\frac{(2n + \beta +2)(n + \beta)(n + 1)}{2n + \beta}} {\phi}_{n+1} (y), 
\end{equation}
 satisfied by the radial functions (\ref{31}). If we introduce the
 number operator defined as
\begin{equation} \label{35}
 \hat{N} \phi_{n}
  = n \phi_{n},
\end{equation}
we can easily find ladder operators \cite{Meljanac:2006uf},\cite{borzov},\cite{Dadic:2002qn}
 from the recursive relations (\ref{34a}) and(\ref{34b}). 
They are
\begin{equation} \label{36}
 b = \left [\hat{N} - \frac{1}{2} y - y \frac{d}{dy}) \right ]  
  \sqrt{\frac{(2\hat{N} + \beta)}{(\hat{N} + \beta -1)(2\hat{N} + \beta -2)}},
\end{equation}
\begin{equation} \label{37}
  b^{\dagger} = \left [\hat{N} + \beta - \frac{1}{2} y + y \frac{d}{dy}) \right ]  
  \sqrt{\frac{(2\hat{N} + \beta)}{(\hat{N} + \beta)(2\hat{N} + \beta +2)}}, 
\end{equation}
Straightforward calculation shows that the ladder operators
$ \; b \; $ and $ \; b^{\dagger} \; $ are bosonic,
\begin{equation} \label{bosonic}
 [b, b^{\dagger}] = 1,
\end{equation}
together with
\begin{equation} \label{bosonicnumber}
 [\hat{N}, b ] = - b,   \;\;\;\;\; [\hat{N}, b^{\dagger} ] = b^{\dagger},
\end{equation}
resulting in the simple relation including the number operator $ \; \hat{N}
= b^{\dagger} b. \; $
If we take the vacuum state to be the wavefunction (\ref{31}) with the
lowest possible energy (which happens when $ n = 0$), it is seen that
$\; b \; $ annihilates this vacuum state, namely, $\; b {\phi}_{0} = 0. \;$
The bosonic operators $\; b \; $ and $\; b^{\dagger} \; $ satisfy the
usual oscillator relations,
\begin{equation} \label{bosonicrel}
 b \phi_{n} = \sqrt{ n} \phi_{n-1},   \;\;\;\;\; 
 b^{\dagger} \phi_{n} = \sqrt{ n+1} \phi_{n+1}.
\end{equation}

Due to relations (\ref{1}),(\ref{32}) and (\ref{35}), the radial Hamiltonian
in Eq. (\ref{radialhamiltonianfirst})
can be expressed in terms of the number operator $\; \hat{N} \;$ in
the way
\begin{equation} \label{38}
 H_{r} = - \frac{1}{4N} \frac{{\alpha}^2}{{(\hat{N} + k + b  + \frac{1}{2})}^2}.
\end{equation}
We would like to factorize the radial Hamiltonian (\ref{38}), so that
 it can be written in the form
\begin{equation} \label{39}
 H_{r} = \frac{1}{{\gamma}^{2}} A^{\dagger} A + h,
\end{equation}
with $\; \gamma \;$ and $\; h \;$ as yet undetermined parameters. 
In order to make such factorization, we have to
 make a transition from the pair of bosonic oscillators $\; \{ b, b^{\dagger} \} \;$ 
to new pair of operators $\; \{ A, A^{\dagger} \}. \;$ Although the
form of the Hamiltonian (\ref{39}) looks much more simple,
the price we have payed is that the new deformed oscillators are no
 more bosonic, but rather they are deformed, obeying a more
 complicated relation in terms of the number operator, namely,
\begin{equation} \label{40}
  A^{\dagger} A = \Phi (\hat{N}).
\end{equation}
In the above relation $\; \Phi (\hat{N})  \;$ is an analytic function
which is required to satisfy following three conditions:
\begin{equation} \label{41}
 (i) \quad \Phi (\hat{N}) > 0,
\end{equation}
\begin{equation} \label{42}
  (ii)  \quad \Phi (0) = 0,  
\end{equation}
\begin{equation} \label{43}
 (iii)  \quad  \Phi (1) = 1.
\end{equation}
The function $\; \Phi (\hat{N})  \;$ which is consistent with  
relations (\ref{39}) and (\ref{40}) and which obeys conditions
(i),(ii) and (iii) is given as
\begin{equation} \label{44}
  \Phi (\hat{N}) =\frac{{\gamma}^{2} {\alpha}^{2}}{4N}
  \left ( \frac{1}{{(\frac{\beta}{2})}^{2}} - \frac{1}{{ (\hat{N} +
  \frac{\beta}{2})}^{2}} \right ).
\end{equation}
The parameters $\; h \; $ and $\; \gamma \; $ are introduced so as to accommodate for
the condition (ii) and the normalization condition (iii), respectively, and are equal to
\begin{equation} \label{45}
  h = - \frac{{\alpha}^{2}}{N {\beta}^{2}}, \quad
 {\gamma}^{2} = \frac{{\beta}^{2} {(\frac{\beta}{2} +
 1)}^{2} N}{{\alpha}^{2} (\beta + 1)}.
\end{equation}
The deformed oscillators $\; A, A^{\dagger} \;$ can be
related \cite{mmp} to the bosonic oscillators (\ref{36}) and (\ref{37})
in the following way
\begin{equation} \label{46}
 A = b ~ \sqrt{\frac{\Phi (\hat{N})}{\hat{N}}}, \quad 
  A^{\dagger} = \sqrt{\frac{\Phi (\hat{N})}{\hat{N}}} ~ b^{\dagger}.
\end{equation}
If we further introduce the operators $ \; J_{+}, J_{-}, J_{0} \; $ defined
as
\begin{equation} \label{47}
  J_{-} = A \frac{\hat{N}}{ \sqrt{\Phi (\hat{N})}}, \quad
 J_{+} = \frac{\hat{N}}{ \sqrt{\Phi (\hat{N})}} A^{\dagger}, \quad
  J_{0} = \hat{N} + \frac{1}{2},
\end{equation}
one can show that they are, in fact, generators of  $\; SU(1,1) \; $ algebra,
\begin{equation} \label{48}
  [J_{-}, J_{+}] = 2 J_{0}, \quad  [J_{0}, J_{\pm}] = \pm J_{\pm}.
\end{equation}

   In papers \cite{ghosh}, \cite{cordero} the
  underlying conformal symmetry of the rational Calogero model with a
  Coulomb-like term is revealed by constructing an explicit
  realizations of the corresponding $ \; SU(1,1) \; $
  generators. These realizations happen to be different from those
  found in \cite{Meljanac:2003jj},\cite{Meljanac:2006uf} where the
  realizations of $ \; SU(1,1) \; $ generators for the rational Calogero model with the
  harmonic confining term are considered.
   Since it is known  \cite{Meljanac:2003jj},\cite{Meljanac:2006uf} that
  all models with underlying conformal symmetry 
  can be mapped to the set of decoupled oscillators, one could do the
  same for  the Hamiltonian (\ref{hamiltonian}) by
  using the construction of $ \; SU(1,1) \; $ generators  made in \cite{ghosh}, \cite{cordero}.
   After finding an appropriate similarity transformation, one could
  apply it to $ \; SU(1,1) \; $ generators to find ladder operators
  for the Hamiltonian (\ref{hamiltonian}). It is possible to carry
  out such transformation since the all systems with underlying conformal
  symmetry  have radial excitations described by the associated Laguerre polynomials
  with two of the generators playing the role of creation and annihilation
  operators in the equivalent problem including decoupled set of oscillators.
   This approach would lead
  to ladder operators which would not coincide with the ladder operators
  (\ref{47}), but would rather be related to them by means of some
  similarity transformation.

\section{Deficiency indices of the radial Hamiltonian}

The spectrum of this model discussed above is valid for the usual boundary 
conditions where the wave function vanishes at $r=0$ and it is square 
integrable. We shall now find the most general set
of boundary conditions for which the radial Hamiltonian $H_r$ is self-adjoint. 
For this we follow the method of von Neumann. We start by recalling the 
essential features of this method \cite{reed}.

Let $T$ be an unbounded differential operator acting on a Hilbert
space ${\cal H}$ and let $D(T)$ be the domain of $T$. The inner product 
of two element $\alpha , \beta \in {\cal H}$ is denoted by $(\alpha ,
\beta)$. Let $D(T^*)$ be the set
of $\phi \in {\cal H}$ for which there is a unique $\eta \in {\cal H}$ with
$(T \xi , \phi) = (\xi , \eta )~ \forall~ \xi \in D(T)$. For each such
$\phi \in D(T^*)$, we define $T^* \phi = \eta$. $T^*$ then defines the adjoint
of the operator $T$ and $D(T^*)$ is the corresponding domain of the adjoint.
The operator $T$ is called symmetric or Hermitian iff $(T \phi, \eta) = 
(\phi, T \eta) ~ \forall ~ \phi, \eta \in D(T)$. The operator $T$ is called 
self-adjoint iff $T = T^*$ {\it and} $D(T) = D(T^*)$. 

We now state the criterion to determine if a symmetric operator $T$ is
self-adjoint. For this purpose let us define the deficiency subspaces 
$K_{\pm} \equiv {\rm Ker}(i \mp T^*)$ and the 
deficiency indices $n_{\pm}(T) \equiv
{\rm dim} [K_{\pm}]$. Then $T$ falls in one of the following categories:\\
1) $T$ is (essentially) self-adjoint iff
$( n_+ , n_- ) = (0,0)$.\\
2) $T$ has self-adjoint extensions iff $n_+ = n_-$. There is a one-to-one
correspondence between self-adjoint extensions of $T$ and unitary maps
from $K_+$ into $K_-$. \\
3) If $n_+ \neq n_-$, then $T$ has no
self-adjoint extensions.

We now return to the discussion of the effective Hamiltonian ${H_r}$.
This is an unbounded differential operator defined in    
$R^+ $. ${H_r}$ is a symmetric operator on the domain
$$
D({H_r}) \equiv \{\phi (0) = \phi^{\prime} (0) = 0,~
\phi,~ \phi^{\prime}~  {\rm absolutely~ continuous},~ 
        \phi \in {\rm L}^2(d\sigma)\}, $$
        where $d\sigma=r^\beta dr$. 
We would next like to determine if ${H_r}$ is self-adjoint in the domain 
$D(H_r)$.
To perform such an analysis it is
necessary to obtain the square-integrable solutions of the equation
\begin{equation} \label{b5}
 H_{r}^{\ast} {\phi}_{\pm} (r) = \pm i {\phi}_{\pm} (r).
\end{equation}
The operator $\; H_{r}^{\ast} \; $ is the adjoint of $\; H_{r} \; $ and
is given by the same differential operator as $\; H_{r}, \; $ although
their domains might be different. Below we shall give the analysis for
the parameter
 range where $\mu \neq 0$, the case for $\mu=0$ being similar.
Thus, Eq.(\ref{b5}) is identical to Eq.(\ref{1}) when $\; \tilde{E} = \pm i, $
\begin{equation} \label{b6}
 -\frac{d^2}{d r^2} {\phi}_{\pm} (r) - (2k + 2b + 1) \frac{1}{r} \frac{d}{d r}
 {\phi}_{\pm} (r) - \frac{\alpha}{\sqrt{N} r} {\phi}_{\pm} (r) = \pm i {\phi}_{\pm} (r).
\end{equation}
We are interested in finding the square-integrable solutions to
 Eq.(\ref{b6}). The solutions
 of Eq.(\ref{b5}) or Eq.(\ref{b6}) which are square-integrable at infinity are given by
 $ \; {\phi}_{\pm} (r) = r^{- \frac{\beta}{2}} {\psi}_{\pm} (r), \; $ where
\begin{equation} \label{b7}
 {\psi}_{\pm} (r) = e^{-\frac{1}{2}c_{\pm} r} {(c_{\pm} r)}^{\frac{\beta}{2}}
   U ( \frac{\beta}{2} - {\kappa}_{\pm},~\beta, ~ c_{\pm} r)
\end{equation}
with
$$
  c_{+} = c(\tilde{E} = i) = 2 \sqrt{-i}, \quad c_{-} =
 c(\tilde{E} = -i) = 2 \sqrt{i},  
$$
\begin{equation} \label{bt4}
 \quad {\kappa}_{+} = \frac{\alpha}{c_{+} \sqrt{N}} = \frac{\alpha}{ \sqrt{-4N i }}, 
    \quad {\kappa}_{-} = \frac{\alpha}{c_{-} \sqrt{N}} = \frac{\alpha}{ \sqrt{4N i }}.
\end{equation}
Since these solutions are also required to be square-integrable near
the origin, it is necessary to investigate their behaviour for
$ \; r \rightarrow 0, \; $
which looks as {\footnote{In subsequent considerations we shall work
    on the space of $\; \psi(r) \; $ functions where the measure is $\;
     dr. $ }}
\begin{equation} \label{21}
 {\psi}_{\pm} (r) \longrightarrow
  {(c_{\pm} r)}^{\frac{\beta}{2}} \frac{\pi}{\sin \pi \beta} 
   \bigg [ \frac{1}{ \Gamma (1- \frac{\beta}{2} - {\kappa}_{\pm})
  \Gamma (\beta)} 
 - \frac{{(c_{\pm} r)}^{1- \beta}}
   { \Gamma (\frac{\beta}{2} - {\kappa}_{\pm})
  \Gamma (2 - \beta)}    \bigg ].
\end{equation}
In the above expression we have restricted ourselves to the lowest
order in $ \; r, \; $ so that we could take 
$ \;  M (a,b,z) \rightarrow 1 \; $
as the argument $ \; z \: $ tends to $ \; 0. $
The square integrability of the wavefunction (\ref{b7}) near the
origin is determined by (\ref{21}) which implies that as 
$\; r \rightarrow 0, \; $
\begin{equation} \label{b8}
 {|{\psi}_{\pm} (r)|}^{2} dr  \longrightarrow 
   \bigg [ A_{1} {r}^{\beta} +  A_{2} r + A_{3} {r}^{2 - \beta}  \bigg ] dr,
\end{equation}
where $ \; A_{1}, A_{2}, A_{3} \: $ are some constants independent of $ \; r. $
From Eq.(\ref{b8}) it is seen that near the origin, the functions $ \; {\psi}_{\pm} \:$ 
(and consequently functions $ \; {\phi}_{\pm} \:$) are not
square-integrable for the parameter $ \; \beta \: $ satisfying
 $ \;\beta < -1 \: $ or  $ \;\beta > 3. \: $ Consequently, in  the
 parameter range $ \;\beta < -1 \: $ or  $ \;\beta > 3, \: $ the
 functions $ \; {\psi}_{\pm} \:$
are not the elements of the vector space  $ \; L^{2}[R^{+}, dr] \:$  of quadratically integrable
functions defined on the positive real axis. In that case,
  $ \; n_{+} = n_{-} = 0 \:$ and   $\; H_{r} \; $ is essentially
  self-adjoint in the domain $ \; {\mathcal{D}} (H_{r}). \;$
However, if $ \; -1 < \beta < 3, \: $ the functions $ \; {\psi}_{\pm} \:$ 
(and consequently functions $ \; {\phi}_{\pm} \:$) are
square-integrable. Thus, if $ \; \beta \:$ lies in this range, we have  
$ \; n_{+} = n_{-} = 1 \:$ and the Hamiltonian  $\; H_{r} \; $ is not
  self-adjoint in the domain $ \; {\mathcal{D}} (H_{r}), \;$ but
  admits self-adjoint extensions. Note that from (\ref{4}), the allowed range of $\beta$ implies that the parameter $\mu$ 
must lie in the range $ -1 < \mu < 1$.

The above allowed range of $\mu$, together with (\ref{parameterb}),
 (\ref{2}) and (\ref{4}), 
 implies that the values of $N$, $k$ and 
$a +\frac{1}{2}$ must satisfy the relation
\be \label{pp}
- \frac{ N - 1 + 2 k}{N ( N-1)} < a + \frac{1}{2} < -\frac{ N - 5 + 2 k}{N ( N-1)}.
\ee
for the self-adjoint extension to exist. 
For $N \geq 3$, we have the following 
classifications of the boundary conditions depending on the value of the
parameter $a+ \frac{1}{2}$.\\
(i) $a + \frac{1}{2}\geq \frac{1}{2}$ : This corresponds to the boundary 
condition considered  by Khare in \cite{khare}. For this choice, 
 both the wave-function and the current vanish as $x_i \rightarrow x_j$.
In this case, $\mu > 1$ for all values of $k \geq 0$.
The corresponding Hamiltonian is essentially self-adjoint in the domain
$D(H_r)$, leading to a unique quantum theory.\\
(ii) $ 0 < a + \frac{1}{2} <  \frac{1}{2}$ :  
For this choice we see that the wave-function in (\ref{factorization}) vanishes in the limit $x_i \rightarrow x_j$, 
although the current may be divergent. In this case 
$\mu > 0$  and $k$ must be equal to zero so that $\mu$ may belong to 
the range $0 < \mu < 1$. The corresponding constraint on $a + \frac{1}{2}$ is given by
$0 < a + \frac{1}{2} < \frac{5 - N }{N(N-1)}$, which can only be
satisfied for $N = 3$ and $4$.  So new quantum states associated with 
the self-adjoint extension of $H_r$ exist only in the $k=0$ sector 
of $N = 3$ and $N=4$.  
\\
(iii) $ -\frac{1}{2} < a + \frac{1}{2} <  0$ :
The lower bound on $a+ \frac{1}{2}$ is obtained from the condition that the
wavefunction be square-integrable. The parameter $a + \frac{1}{2} $ in this
range leads to a singularity in the wavefunction $\Psi$ in
Eq. (\ref{factorization})
resulting from the coincidence of any two or more particles. 
Using permutation symmetry, such an eigenfunction can be extended to the
whole of configuration space, although not in a smooth fashion.
The new quantum states in this case exist for arbitrary $N$ and even for
non-zero values of $k$. In fact,
imposing the condition that the upper bound on $a + \frac{1}{2}$  should be
greater than $-\frac{1}{2}$, we find from  (\ref{pp}) that $k$ is restricted as 
$k < \frac{1}{4} \left ( N^2 - 3 N + 10 \right )$. It can also be
shown that there are only two allowed values of $k$ when both $N$ and 
$a + \frac{1}{2}$ are kept fixed.

Von Neumann's method also provides a prescription for obtaining the
 domain of self-adjointness
 of a symmetric operator, which admits a self-adjoint extension.
The extended domain 
$ \; {\mathcal{D}}_{z} (H_{r}) \;$ in which $\; H_{r} \; $ is
self-adjoint contains all the elements of $ \; {\mathcal{D}} (H_{r}), \;$
together with the elements of the form 
 $ \; e^{i \frac{z}{2}} {\psi}_{+} + 
 e^{-i \frac{z}{2}} {\psi}_{-}, \;$ where $ \; z \in R ~ (mod ~ 2
 \pi ). \:$ Thus the self-adjoint extensions of this model exist when 
 $ \; -1 < \beta < 3, \: $ and in that case,
 $$ \; {\mathcal{D}}_{z}(H_{r}) = {\mathcal{D}}(H_{r})
\oplus  \{e^{i \frac{z}{2}} {\psi}_{+} + e^{-i \frac{z}{2}} {\psi}_{-}
\} \; $$ is the extended domain in which $\; H_{r} \; $
is self-adjoint.


\section{Bound states of the radial Hamiltonian with self-adjoint extension}
We shall now find solutions of the physical problem for the range of system
parameters where the self-adjoint extension is necessary.  
In finding the solutions to Eq.(\ref{1}), we shall first
  consider the bound state sector of the problem. In this sector, the
  energy of the system is negative, $\; E < 0, \;$ and the
  wavefunctions need to be square-integrable. We consider the solution of the form
\begin{equation} \label{b1}
   \psi(r) = B e^{-\frac{cr}{2}} {(cr)}^{\frac{\beta}{2}}
    U \left ( \frac{\beta}{2} - \kappa,~ \beta,~ c r \right ).
\end{equation}
In order to make an analysis and to find the spectrum, we should know
the behaviour of the $ \; U \; $ function near the origin. 
Using Eqs. (\ref{7}) and (\ref{8}), we can expand  $U (a,b,z)$ at 
$ \; z \rightarrow 0$ limit as 
\begin{eqnarray} 
   U (a,b,z) \longrightarrow  \frac{\pi}{\sin \pi b} \bigg [
  \frac{1}{\Gamma (1+a-b) \Gamma(b)} \left [ 1 + \frac{a}{b} z +
   \frac{a(a+1)}{b(b+1)} \frac{z^2}{2!}+ O(z^3) \right ] - \nonumber \\
  ~ \frac{z^{1-b}}{\Gamma(a)
   \Gamma (2-b) } \left [ 1 + \frac{1+a-b}{2-b} z +
   \frac{(1+a-b)(2+a-b)}{(2-b)(3-b)} \frac{z^2}{2!}+ O(z^3) \right ] 
   \bigg ]. \label{b2}
\end{eqnarray}
Consequently,  at $ \; r \rightarrow 0 $ limit, 
$\psi(r)$ in Eq. (\ref{b1}) behaves as
\begin{eqnarray} 
 \psi (r) \longrightarrow B
  e^{-\frac{1}{2}c r} {(c r)}^{\frac{\beta}{2}} \frac{\pi}{\sin \pi \beta} 
   \bigg [ \frac{1}{ \Gamma (1- \frac{\beta}{2} - \kappa )
  \Gamma (\beta)} 
  \left ( 1 + \frac{ \frac{\beta}{2} - \kappa}{\beta} c
  r + \frac{ (\frac{\beta}{2} - \kappa) ( \frac{\beta}{2} -
  \kappa +1)}{2\beta (\beta +1) } {(c r)}^{2} + O(r^3)  \right ) - 
  \nonumber \\
   \frac{{(c r)}^{1- \beta}}
   { \Gamma (\frac{\beta}{2} - \kappa )
  \Gamma (2 - \beta)} 
 \left ( 1 + \frac{ 1- \frac{\beta}{2} - \kappa}{2 - \beta} c
  r + \frac{ (1 -\frac{\beta}{2} - \kappa)
 ( 2 -\frac{\beta}{2} - \kappa)}{2 (2 - \beta) (3- \beta)}
  {(c r)}^{2} + O(r^3)  \right )   \bigg ].
  \label{b3}
\end{eqnarray}
The parameters $\; c \;$ and $\; \kappa \;$
appearing in (\ref{b3}) are given in (\ref{4}), except
that they are evaluated for $\; \tilde{E} = E. \;$ Since the energy is
negative, these parameters are real,
$$
 c = 2 \sqrt{-E} = 2 \sqrt{E_{b}} \equiv p,
$$
\begin{equation} \label{b4}
 \kappa = \frac{\alpha}{c \sqrt{N}} = \frac{\alpha}{ \sqrt{-4N E }} = 
   \frac{\alpha}{ \sqrt{4N E_{b} }} = \frac{\alpha}{p \sqrt{N}}. 
\end{equation}
 Here, for future convenience, we have introduced the absolute value
 $ \; E_{b}  \; $
 of the bound state energy,  $ \; E_{b} = -E, \; E < 0,  \; $ and the
real  parameter $\; p \: $ which coincides with $\; c \; $ in the
 bound state sector.
Note that $\; c \;$ and $\; \kappa \;$ will no more be real in the
scattering sector.

 If the wavefunction (\ref{b1}) is expected to describe a physically
 acceptable bound state solutions to Eq.(\ref{1}), it has to belong
 to the domain of self-adjointness  $ \; {\mathcal{D}}_{z}(H_{r}). \; $
 If $\psi_0 (r) \in D (H_r)$, then an arbitrary element of the domain 
 $ \; {\mathcal{D}}_{z}(H_{r}) \; $ can be written as
$ \psi_0(r) + \rho (e^{i \frac{z}{2}} {\psi}_{+} + 
e^{-i \frac{z}{2}} {\psi}_{-}), $ 
where $ \; \rho \: $ is a constant. 
If the solution of the physical wavefunction (\ref{b1})  belongs to the domain 
$ \; {\mathcal{D}}_{z}(H_{r}) \; $, 
the functional form of physical wavefunction must match 
with that of an arbitrary element of the domain $ \; {\mathcal{D}}_{z}(H_{r})\; $, which is given by
\begin{equation} \label{b9}
\psi (r)=
\psi_0(r) + \rho (e^{i \frac{z}{2}} {\psi}_{+} + e^{-i \frac{z}{2}} {\psi}_{-}),
\end{equation}
Inserting Eqs.(\ref{b3}) and (\ref{21})
into relation (\ref{b9}), and equating the coefficients of the lowest order
powers in $ \; r\; $ (for which there is no contribution 
from $\psi_0(r)$),  yields the following two conditions
$$
 \frac{
 \tilde{B} c^{\frac{\beta}{2}}}{\Gamma (1- \kappa - \frac{\beta}{2})}
    = \frac{e^{i \frac{z}{2}} {c_{+}}^{\frac{\beta}{2}}}{\Gamma (1- {\kappa}_{+} - \frac{\beta}{2})}
  + \frac{e^{-i \frac{z}{2}} {c_{-}}^{\frac{\beta}{2}}}{\Gamma (1- {\kappa}_{-} - \frac{\beta}{2})},
$$
\begin{equation} \label{b10}
 \frac{\tilde {B} c^{1 - \frac{\beta}{2}}}{\Gamma (\frac{\beta}{2} - \kappa)}
    = \frac{e^{i \frac{z}{2}} {c_{+}}^{1- \frac{\beta}{2}}}{\Gamma
    (\frac{\beta}{2} - {\kappa}_{+})}
  + \frac{e^{-i \frac{z}{2}} {c_{-}}^{1 - \frac{\beta}{2}}}{\Gamma (
    \frac{\beta}{2} - {\kappa}_{-})}, 
\end{equation}
where $\tilde {B}= B/\rho $.
After dividing both sides of these two expressions, we get the
relation
\begin{equation} \label{b11}
  \frac{\Gamma (1- \kappa - \frac{\beta}{2})}{\Gamma (\frac{\beta}{2} - \kappa)}
         c^{1 - \beta} =
   \frac{\frac{e^{i \frac{z}{2}} {c_{+}}^{1- \frac{\beta}{2}}}{\Gamma
    (\frac{\beta}{2} - {\kappa}_{+})}
  + \frac{e^{-i \frac{z}{2}} {c_{-}}^{1 - \frac{\beta}{2}}}{\Gamma (
    \frac{\beta}{2} - {\kappa}_{-})}}
   {\frac{e^{i \frac{z}{2}} {c_{+}}^{\frac{\beta}{2}}}{\Gamma (1- {\kappa}_{+} - \frac{\beta}{2})}
  + \frac{e^{-i \frac{z}{2}} {c_{-}}^{\frac{\beta}{2}}}{\Gamma (1-
         {\kappa}_{-} - \frac{\beta}{2})}}.
\end{equation}
Inserting the expressions (\ref{b4}) for $\; c \; $ and $\; \kappa \; $
we obtain the final condition
\begin{equation} \label{b12}
  \frac{\Gamma (1- \frac{\beta}{2} - \frac{\alpha}{\sqrt{4N E_{b}}})}
     {\Gamma (\frac{\beta}{2} - \frac{\alpha}{\sqrt{4N E_{b}}})}
         {(2 \sqrt{E_{b}})}^{1 - \beta} =
   \frac{\frac{e^{i \frac{z}{2}} {c_{+}}^{1- \frac{\beta}{2}}}{\Gamma
    (\frac{\beta}{2} - {\kappa}_{+})}
  + \frac{e^{-i \frac{z}{2}} {c_{-}}^{1 - \frac{\beta}{2}}}{\Gamma (
    \frac{\beta}{2} - {\kappa}_{-})}}
   {\frac{e^{i \frac{z}{2}} {c_{+}}^{\frac{\beta}{2}}}{\Gamma (1- {\kappa}_{+} - \frac{\beta}{2})}
  + \frac{e^{-i \frac{z}{2}} {c_{-}}^{\frac{\beta}{2}}}{\Gamma (1-
         {\kappa}_{-} - \frac{\beta}{2})}},
\end{equation}
which determines the spectrum corresponding to bound states of the
radial Hamiltonian $H_r$ (\ref{radialhamiltonianfirst}) as well as the 
initial many-body Hamiltonian (\ref{hamiltonian}).

Writing $\frac{c_+^{1 - \frac{\beta}{2}}}{\Gamma (\frac{\beta}{2} - \kappa_+)} = \xi_1 e^{i \theta_1}$ and 
$\frac{c_+^{\frac{\beta}{2}}}{\Gamma (1 -  \kappa_+ - \frac{\beta}{2}) } = \xi_2 e^{i \theta_2}$, (\ref{b12}) can 
be expressed as
\begin{equation} \label{usual}
\frac{\Gamma (1- \frac{\beta}{2} - \frac{\alpha}{\sqrt{4N E_{b}}})}
     {\Gamma (\frac{\beta}{2} - \frac{\alpha}{\sqrt{4N E_{b}}})}
         {(2 \sqrt{E_{b}})}^{1 - \beta} =
\frac{ \xi_1 cos ( \theta_1 + \frac{z}{2})}{ \xi_1 cos ( \theta_2 + \frac{z}{2})}.
\end{equation}

The above analysis shows that for a given choice of the system parameters, Eq. (\ref{usual}) gives the
energy eigenvalue $E = -E_b$ as a function of the self-adjoint extension parameter
$z$. For a fixed set of system parameters, different choices of $z$
lead to inequivalent quantization and to the spectrum for this model
in the parameter range where the system admits
 self-adjoint extension. In general, the energy $E = -E_b$ cannot be
 calculated analytically and
 has to be obtained numerically by plotting (\ref{usual}). Figures 1
 and 2 show l.h.s and r.h.s of Eq.(\ref{usual}) for two different,
 representative sets of the system parameters as well as for the two
 different choices of the self-adjoint extension parameter
 $z$. The curved lines at those figures  represent graph of the function $f(E_b) $ which is given by the
 l.h.s of Eq.(\ref{usual}). On the other hand
 r.h.s of Eq.(\ref{usual}) is represented by a horizontal
 straight line. The energy
 eigenvalues of the system described by the Hamiltonian
 (\ref{hamiltonian}) are obtained by looking at the  intersections of
 these two curves.
 We see from figures that there is an infinite number of bound states near $E_b \rightarrow 0$.
For $\; \alpha > 0, \; $ there are infinite number of bound
  states for any value of $z$. However, the existence of
  non-oscillatory part shows that, just like the usual case, the
  spectrum has a lower bound for all possible values of $z$.
  The situation when $\; \alpha > 0 \; $ is shown at figures 1 and 2.  

For the choice of the self-adjoint extension parameter $z = z_1$ such
 that $ \theta_1 + \frac{z_1}{2} = \frac{\pi}{2}$, the r.h.s. of 
(\ref{usual}) is zero. This implies that 
\begin{equation} \label{usual1}
\frac{\beta}{2} - \frac{\alpha}{\sqrt{4N E_{b}}} = -n, ~~~~~ n = 0,1,2,....
\end{equation}
which gives the usual energy eigenvalues as expressed in
(\ref{32}). It can be shown
 that the choice of $z = z_2$ such that 
$ \theta_2 + \frac{z_2}{2} = \frac{\pi}{2}$ gives a similar result.
 At this point it may be noted that the analytical solution
 (\ref{usual1}) implies that for a certain values of the self-adjoint
 extension parameter and system parameters, even the repulsive Coulomb potential leads
 to the formation of only one bound state.
It can easily be seen if we write (\ref{usual1}) in the form
$ \; \alpha = \sqrt{N E_b}(2n + \beta). \; $ This expression shows that
in order to have the repulsive Coulomb potential, that is $\; \alpha < 0, \; $
 one has to restrict $\; \beta \; $ within the range
$\; -1 < \beta < 0 \; $ and set $n$ equal to zero, resulting in a
single bound state.
The same conclusion holds also in the general case where the analytical
solution is not possible, and it can be verified by extensive numerical
investigation of the general relation (\ref{b12}) (see Figure 3 as an example).

\begin{figure}
\begin{center}
\includegraphics[width=9cm]{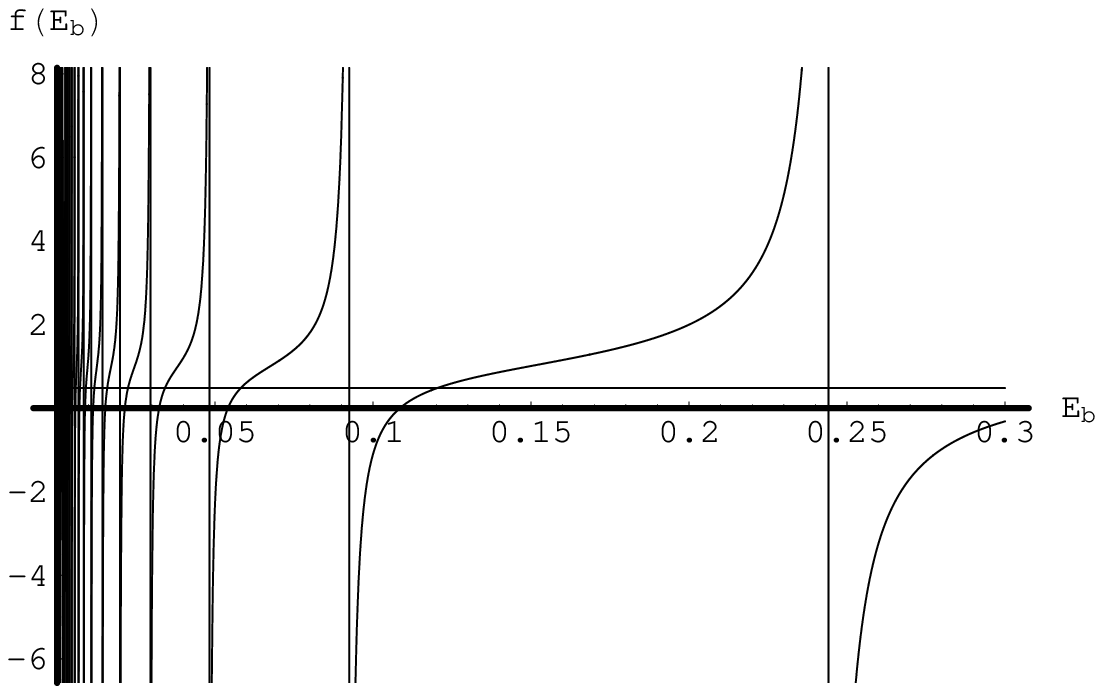}
\end{center}
\noindent
\begin{flushleft}
Figure 1. A plot of Eq. (\ref{usual}) using Mathematica
with $N = 1000$, $\alpha = 50$,  $\beta = 0.8 $, $ k = 1 $ and 
$z = 0.1$. The horizontal straight line corresponds to the value of the r.h.s of
Eq.(\ref{usual}).
\end{flushleft}
\end{figure}

\begin{figure}
\begin{center}
\includegraphics[width=9cm]{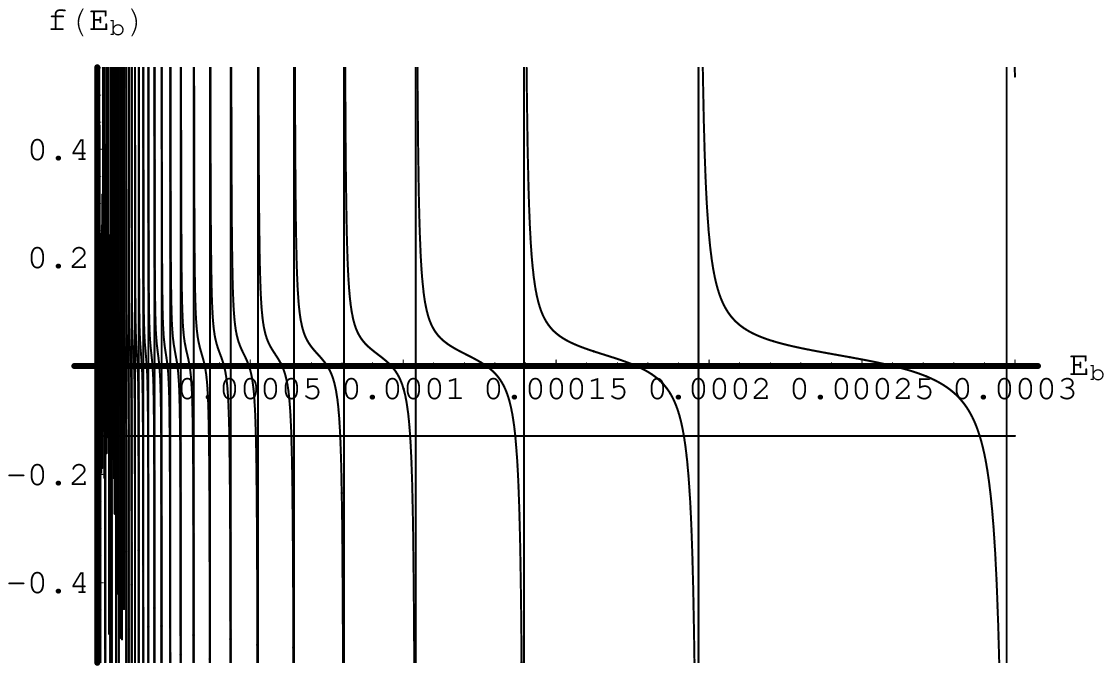}
\end{center}
\noindent
\begin{flushleft}
Figure 2. A plot of Eq. (\ref{usual}) using Mathematica
with $N = 100$, $\alpha = 1.5$,  $\beta = -0.7 $, $ k = 1 $ and 
$z = -0.73$. The horizontal straight line corresponds to the value of the r.h.s of
Eq.(\ref{usual}).
\end{flushleft}
\end{figure}

\begin{figure}
\begin{center}
\includegraphics[width=9cm]{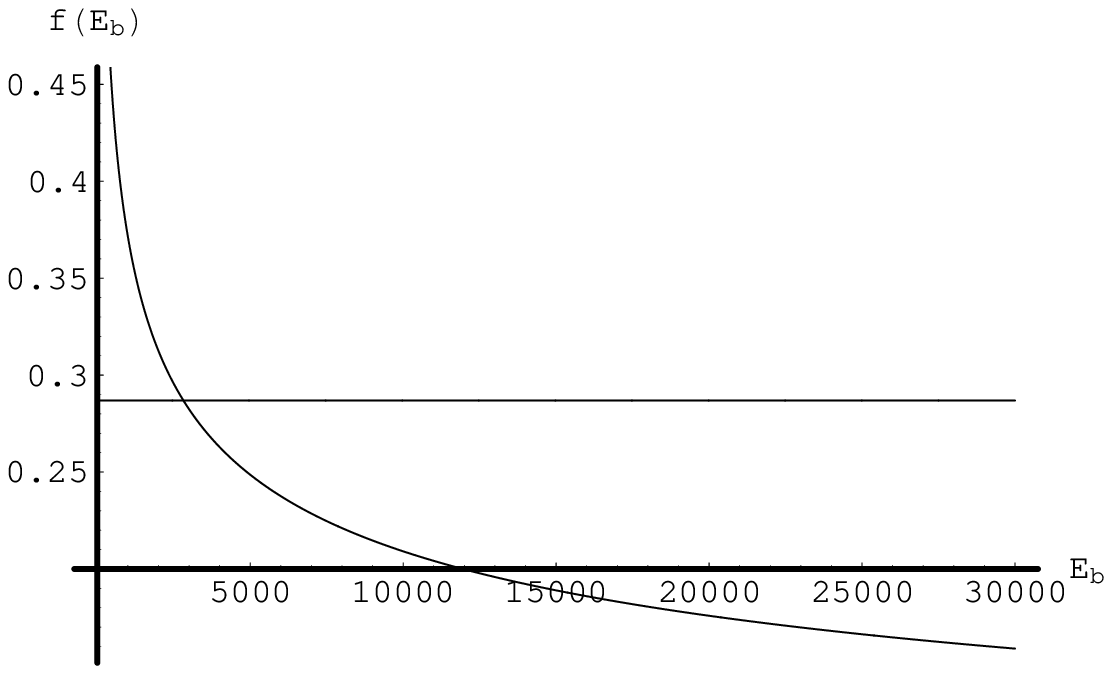}
\end{center}
\noindent
\begin{flushleft}
Figure 3. A plot of Eq. (\ref{usual}) using Mathematica
with $N = 1000$, $\alpha = -1$,  $\beta = 1.5 $, $ k = 1 $ and 
$z = 0.1$. The horizontal straight line corresponds to the value of the r.h.s of
Eq.(\ref{usual}). This graph shows the general feature exhibited for
arbitrarily strong repulsive Coulomb potential, i.e. for any $\; \alpha < 0.$
\end{flushleft}
\end{figure}

Finally, it is important to emphasize that the
       system described by (\ref{hamiltonian}) has a fundamentally different behaviour depending on the sign of $\alpha$.
       While for $\; \alpha > 0, \;$ the
       l.h.s. of Eq.(\ref{usual}) exhibits oscillatory, as well as
       non-oscillatory behaviour, leading to
       infinite number of bound states,
       for $\; \alpha \leq 0, \;$ it shows only non-oscillatory
       behaviour resulting in the existence of at most one bound state.
       This single bound state, if it exists, shows up only for the certain range of the
       self-adjoint extension parameter $z$.  For $\; \alpha = 0, \; $ this
   observation is consistent with the result obtained in \cite{npb}.
   This feature can most easily be seen by looking at the special case
  (58) where the analytical solution is available. There, in order for
  $\; \alpha \; $ to be less than zero, we must have $\; n = 0 \; $
  together with $\; \beta \; $ within the range
    $\; -1 < \beta < 0, \; $ resulting in a single bound state 
   $\; E = -E_b = - \frac{{\alpha}^{2}}{N {\beta}^{2}}, \; $ as
  already stated just after Eq.(58).


\section{Scattering states of the radial Hamiltonian with self-adjoint 
   extension}
Let us now turn our attention to the scattering sector of the problem
described by Eq.(\ref{1}).  
Since the scattering states correspond to positive energy solutions of 
Eq.(\ref{1}) when $  \tilde{E} = E>0,$   
the variable $y = cr = 2 \sqrt{- E } r$ 
becomes purely imaginary, i.e.  
$y =    iqr, ~$ where the real parameter $q$ is defined as
$q=2 \sqrt{ E } \;  $. Therefore, for analyzing 
the $r \rightarrow \infty$ limit of the scattering states, 
it is of importance to know the behaviour of the
confluent hypergeometric functions $M (a,b,z)$ and $U (a,b,z)$
in the asymptotic region ${\rm Re} (z)=0$ 
and ${\rm Im} (z) \rightarrow +\infty $. 
 Following Abramowitz \& Stegun, 
we can expand confluent hypergeometric functions in this 
asymptotic region as 
\begin{equation} \label{9}
   M (a,b,z) \longrightarrow 
   \frac{\Gamma (b)}{\Gamma (a)} e^z z^{a-b} \bigg [ 1+
   O({|z|}^{-1}) \bigg ] + \frac{\Gamma (b)}{\Gamma (b - a)} {(-z)}^{-a} \bigg [ 1+
   O({|z|}^{-1}) \bigg ], 
\end{equation}
\begin{equation} \label{10}
   U (a,b,z) \longrightarrow  O({|z|}^{-a}) .
\end{equation}
Due to the fact that we are dealing with the problem where $\; {\rm Re} (z)
= 0, \;$ both leading terms in the asymptotic expansion (\ref{9}) of
$M$ approximately have the contribution of the same order, so that
both of them have to be taken into account.
In order to find the scattering matrix we could equally well take the
 following linear combination
\begin{equation} \label{11}
  \chi (y) = A M \left ( \frac{1}{2} + \mu - \kappa,~ 1 + 2\mu,~ y
  \right )
   + B y^{- 2 \mu} M \left ( \frac{1}{2} - \mu - \kappa,~ 1 - 2\mu,~ y \right ),
\end{equation}
as a general solution to Eq.(\ref{3}), instead of the one given in (\ref{5}). In this case the solution for
the function $ \psi, $ appearing in (\ref{2}) would look like
\begin{equation} \label{12}
  \psi(r) = e^{-\frac{1}{2}cr} {(cr)}^{\frac{\beta}{2}}
 \left (   A(q) M \left ( \frac{\beta}{2} - \kappa,~ \beta,~ c r \right )
  + B(q) {(cr)}^{1 - \beta} M \left (1 - \frac{\beta}{2} -
 \kappa,~ 2 - \beta,~ c r \right ) \right ),
\end{equation}
where we have assumed that the coefficients $A(q)$ and
$B(q)$ depend on the real parameter $q$. 

By using (\ref{9}), we have the following 
$ r \rightarrow \infty $ limits
\begin{equation} \label{13}
   M (\frac{\beta}{2} - \kappa,~ \beta,~ c r ) \longrightarrow 
     \frac{\Gamma (\beta)}{\Gamma (\frac{\beta}{2} - \kappa)}
     e^{cr} {(cr)}^{- \frac{\beta}{2} - \kappa}
     + \frac{\Gamma (\beta)}{\Gamma (\frac{\beta}{2} + \kappa)}
      {(-cr)}^{- \frac{\beta}{2} + \kappa},
\end{equation}
\begin{equation} \label{14}
   M (1 - \frac{\beta}{2} - \kappa,~ 2 - \beta,~ c r ) \longrightarrow 
     \frac{\Gamma (2 - \beta)}{\Gamma (1 - \frac{\beta}{2} - \kappa)}
     e^{cr} {(cr)}^{ \frac{\beta}{2} - \kappa -1} +
      \frac{\Gamma (2 - \beta)}{\Gamma (1 - \frac{\beta}{2} + \kappa)}
      {(-cr)}^{ \frac{\beta}{2} + \kappa -1},
\end{equation}
so that the wave function (\ref{12}), describing the scattering state,
in the above limit behaves as
$$
  \psi(r) \equiv \psi( \tilde{E} = E)  \longrightarrow 
 A(q) \frac{\Gamma (\beta)}{\Gamma (\frac{\beta}{2} - \kappa)}
     e^{\frac{1}{2} cr}  {(cr)}^{- \kappa}  +
   A(q) \frac{\Gamma (\beta)}{\Gamma (\frac{\beta}{2} + \kappa)}
      e^{-\frac{1}{2} cr} {(-1)}^{-\frac{\beta}{2} + \kappa} {(cr)}^{ \kappa} 
$$
\begin{equation} \label{15}
 + B(q)  \frac{\Gamma (2 - \beta)}{\Gamma (1 - \frac{\beta}{2} - \kappa)}
   e^{\frac{1}{2} cr} {(cr)}^{- \kappa}
      + B(q) {(-1)}^{\frac{\beta}{2} + \kappa -1} \frac{\Gamma (2 -
     \beta)}{\Gamma (1 - \frac{\beta}{2} + \kappa)}
   e^{-\frac{1}{2} cr} {(cr)}^{ \kappa}.
\end{equation}
Note that the parameter $\; \kappa \;$ is also purely imaginary in the
scattering sector. For the coupling constant $\; \alpha \; $ greater
than zero, $\; \kappa \;$ can be expressed as  $\; \kappa = -i
\frac{\alpha}{q \sqrt{N}} = -i \frac{|\alpha|}{q \sqrt{N}} = - i |\kappa|. \;$
By using the relations $y=cr=iqr$ and $\kappa = - i |\kappa|$, we 
can express  $ \psi(r) $ in Eq. (\ref{15}) in
terms of oscillatory incoming wave and 
outgoing wave as 
$$
  \psi(r) \equiv \psi( \tilde{E} = E)  \longrightarrow 
    e^{-i \frac{\pi}{2} \kappa} q^{- \kappa}
  \left ( A(q) \frac{\Gamma (\beta)}{\Gamma (\frac{\beta}{2} - \kappa)}
    + B(q)  \frac{\Gamma (2 - \beta)}{\Gamma (1 - \frac{\beta}{2} -
  \kappa)} \right ) e^{i ( \frac{1}{2} qr + | \kappa| \ln r )} +
$$
\begin{equation} \label{16}
   + e^{i \frac{\pi}{2} \kappa} q^{ \kappa}
  \left ( A(q) \frac{\Gamma (\beta)}{\Gamma (\frac{\beta}{2} + \kappa)}
      e^{i \pi (\kappa - \frac{\beta}{2})}
   + B(q)  \frac{\Gamma (2 - \beta)}{\Gamma (1 - \frac{\beta}{2} +
  \kappa)}  e^{i \pi (\kappa + \frac{\beta}{2} -1)} \right )
    e^{-i ( \frac{1}{2} qr + | \kappa| \ln r )}. 
\end{equation}
The scattering matrix and the corresponding phase shift can be obtained
 from the above limiting form of the wave function as a ratio of
its outgoing and incoming amplitudes,
\begin{equation} \label{17}
  S (q) =  e^{2i \varphi (q)} 
   = \frac{ \left ( A(q) \frac{\Gamma (\beta)}{\Gamma (\frac{\beta}{2} - \kappa) } 
   + B(q) \frac{\Gamma (2 - \beta)}{\Gamma (1 - \frac{\beta}{2} -
   \kappa)} \right )  
     q^{- 2 \kappa} e^{- i \pi \kappa} }         
   { \left ( A(q) \frac{\Gamma (\beta)}{\Gamma (\frac{\beta}{2} +
   \kappa) } e^{i \pi ( \kappa - \frac{\beta}{2})}  
   + B(q) \frac{\Gamma (2 - \beta)}{\Gamma (1 - \frac{\beta}{2} +
   \kappa)}  e^{i \pi ( \kappa + \frac{\beta}{2} - 1)} \right )}.
\end{equation}

Next, to find a relationship between so far unspecified constants
$A(q)$ and $B(q),$ we use the expansion (\ref{7}) to obtain the $ r \rightarrow 0$  limit
of the wave function (\ref{12}), in the lowest order in $r,$
\begin{equation} \label{18}
  \psi( \tilde{E} = E)  \longrightarrow  A(q) {(cr)}^{\frac{\beta}{2}} +
  B(q) {(cr)}^{1 - \frac{\beta}{2}}.
\end{equation}
We recall that the Hamiltonian $H_{r}$ admits a self-adjoint extension
in the parameter range $ 3 > \beta > -1. $  
Since the wave function (\ref{12}) has to belong to the domain of
self-adjointness  ${\mathcal{D}}_{z}(H_{r}) = {\mathcal{D}}(H_{r})
\oplus  \{e^{i \frac{z}{2}} {\psi}_{+} + e^{-i \frac{z}{2}} {\psi}_{-}
\}$ we can write  
\begin{equation} \label{19}
  \rho \, \psi( \tilde{E} = E) = e^{i \frac{z}{2}} {\psi}_{+} + e^{-i \frac{z}{2}} {\psi}_{-},
\end{equation}
where $\rho$ is some constant and, as before, ${\psi}_{\pm}$ are square integrable
solutions of Eq.(\ref{1}) when $ \tilde{E} = \pm i,$ respectively.
In the limit $\; r \rightarrow 0 ,\; $ the behaviour of ${\psi}_{\pm}$
is given by the relation (\ref{21}).
Since according to Eq.(\ref{19}), the coefficients of appropriate
powers of $r$ in (\ref{18}) and (\ref{21}) must match, the following two
conditions emerge
\begin{equation} \label{22}
  \rho\, A(q) c^{\frac{\beta}{2}} = e^{i \frac{z}{2}} 
   \frac{\pi}{\sin \pi \beta} 
   \frac{{c_{+}}^{\frac{\beta}{2}}}{ \Gamma (1- \frac{\beta}{2} - {\kappa}_{+}) \Gamma (\beta)}
  + e^{-i \frac{z}{2}} 
   \frac{\pi}{\sin \pi \beta} 
   \frac{{c_{-}}^{\frac{\beta}{2}}}{ \Gamma (1- \frac{\beta}{2} - {\kappa}_{-}) \Gamma (\beta)},
\end{equation}
\begin{equation} \label{23}
  \rho \, B(q) c^{1 - \frac{\beta}{2}} = - e^{i \frac{z}{2}} 
   \frac{\pi}{\sin \pi \beta} 
   \frac{{c_{+}}^{1 -\frac{\beta}{2}}}{ \Gamma (\frac{\beta}{2} - {\kappa}_{+}) \Gamma (2 - \beta)}
  - e^{-i \frac{z}{2}} 
   \frac{\pi}{\sin \pi \beta} 
   \frac{{c_{-}}^{1 - \frac{\beta}{2}}}{ \Gamma (\frac{\beta}{2} - {\kappa}_{-}) \Gamma (2 - \beta)}.
\end{equation}
The last two equations yield
\begin{equation} \label{24}
   \frac{A(q)}{B(q)} = - \frac{\Gamma (2 - \beta)}{\Gamma (\beta)}
     \frac{e^{i \frac{z}{2}}  
   \frac{{c_{+}}^{\frac{\beta}{2}}}{ \Gamma (1- \frac{\beta}{2} - {\kappa}_{+})}
  + e^{-i \frac{z}{2}}  
   \frac{{c_{-}}^{\frac{\beta}{2}}}{ \Gamma (1- \frac{\beta}{2} - {\kappa}_{-})}}
   {e^{i \frac{z}{2}}  
   \frac{{c_{+}}^{1 -\frac{\beta}{2}}}{ \Gamma (\frac{\beta}{2} - {\kappa}_{+})}
     + e^{-i \frac{z}{2}} 
   \frac{{c_{-}}^{1 - \frac{\beta}{2}}}{ \Gamma (\frac{\beta}{2} - {\kappa}_{-})}}
          c^{1 - \beta}.
\end{equation}
By using this expression, the scattering matrix (\ref{17}) becomes
\begin{equation} \label{25}
  S (q) =  e^{2i \varphi (q)} 
   = \frac{\frac{F_{2}(\beta,~ \alpha,~ z)}{F_{1}(\beta,~ \alpha,~ z)}
   \frac{e^{i \frac{\pi}{2} (1 - \beta)}  q^{1 - \beta}}{\Gamma (\frac{\beta}{2} - \kappa)} - \frac{1}
{\Gamma (1- \frac{\beta}{2} - \kappa)}}
{\frac{F_{2}(\beta,~ \alpha,~ z)}{F_{1}(\beta,~ \alpha,~ z)}
   \frac{ e^{i \pi (\kappa - \beta + \frac{1}{2})} q^{1 - \beta} }
{\Gamma (\frac{\beta}{2} + \kappa)} - \frac{e^{i \pi (\frac{\beta}{2}
   + \kappa -1)}}{\Gamma (1- \frac{\beta}{2} + \kappa)}}
 e^{-i \pi \kappa} q^{-2\kappa}, 
\end{equation}
where $\; c = 2 \sqrt{-E} = iq, \; $ and
$\; \kappa = \frac{\alpha}{c \sqrt{N}} = \frac{\alpha}{ \sqrt{-4N E }}. \; $  
In writing the expression for the scattering matrix we have introduced
the following two functions
\begin{equation} \label{26}
  F_{1}(\beta,~ \alpha,~ z) = e^{i \frac{z}{2}}  
   \frac{{c_{+}}^{1 -\frac{\beta}{2}}}{ \Gamma (\frac{\beta}{2} - {\kappa}_{+})}
     + e^{-i \frac{z}{2}} 
   \frac{{c_{-}}^{1 - \frac{\beta}{2}}}{ \Gamma (\frac{\beta}{2} - {\kappa}_{-})}, 
\end{equation}
\begin{equation} \label{27}
  F_{2}(\beta,~ \alpha,~ z) = e^{i \frac{z}{2}}  
   \frac{{c_{+}}^{\frac{\beta}{2}}}{ \Gamma (1- \frac{\beta}{2} - {\kappa}_{+})}
  + e^{-i \frac{z}{2}}  
   \frac{{c_{-}}^{\frac{\beta}{2}}}{ \Gamma (1- \frac{\beta}{2} - {\kappa}_{-})},
\end{equation}
where $\; z \; $ is the self-adjoint extension parameter and
 $\; c_{\pm} \; $ and $\; {\kappa}_{\pm}\; $ are defined in (\ref{bt4}).
 As a remark, one
 can note that the functions $\; F_{1} \;$ and $\; F_{2} \;$ are simply related as
$$
  F_{2}(\beta,~ \alpha,~ z) =  F_{1}(2 - \beta,~ \alpha,~ z).
$$

As it is seen from the form of the scattering matrix, for any given
value of $\beta$ in the parameter range which admits self-adjoint
extension, the scattering matrix has an infinite set of poles on the positive imaginary
axis of the complex $q$-plane. The existence of poles for the
scattering matrix means that there are
 bound states in the system under consideration. By taking $\; q = ip
 \;$ as some arbitrary pole for the scattering matrix (\ref{25}), one
 can obtain the following  equation determining the bound state
 energies $\; E_b = -E = \frac{p^2}{4} \;$:
\begin{equation} \label{28}
 \frac{F_{2}(\beta,~ \alpha,~ z)}{F_{1}(\beta,~ \alpha,~ z)}
   \frac{ e^{i \pi (\kappa - \beta + \frac{1}{2})} q^{1 - \beta} }
{\Gamma (\frac{\beta}{2} + \kappa)} - \frac{e^{i \pi (\frac{\beta}{2}
   + \kappa -1)}}{\Gamma (1- \frac{\beta}{2} + \kappa)} = 0. 
\end{equation}
This expression, after utilizing the set of relations $\; q = 2\sqrt{E} = ip = i 2
\sqrt{E_b} \;$ and $\; \kappa = -i \frac{\alpha}{q \sqrt{N}} =
 -\frac{\alpha}{p \sqrt{N}} = -\frac{\alpha}{2\sqrt{E_b} \sqrt{N}},
 \;$ finally gives
\begin{equation} \label{29}
  \frac{\Gamma (1- \frac{\beta}{2} - \frac{\alpha}{2\sqrt{E_{b}} \sqrt{N} })}{\Gamma (\frac{\beta}{2}
   - \frac{\alpha}{2\sqrt{E_b} \sqrt{N}})} p^{1 - \beta}
   =   \frac{F_{1}(\beta,~ \alpha,~ z)}{F_{2}(\beta,~ \alpha,~ z)}, 
\end{equation}
which reproduces the bound state condition (\ref{b11}).

\section{Conclusions}

In this paper we have analyzed the $N$-body rational Calogero model with a Coulomb like interaction.
We have shown that for certain ranges of the system parameters, the system admits a one parameter family 
of self-adjoint extensions. The results obtained here for both bound and scattering state sectors are very 
different from those obtained by Khare in \cite{khare}. However, there is no contradiction between these findings as
they refer to different ranges of the system parameters. We have also shown that for specific choices of the 
self-adjoint extension parameter, the usual results of Khare can be recovered.

It has also been shown that a ladder operator construction exists for this system, which also leads to the 
solution found by Khare. This construction indicates that su(1,1) can be regarded as a spectrum generating algebra 
for this system, as it happens in conformal quantum mechanics
\cite{fub} yielding equispaced energy levels. We think that there is a
strong correlation between our and the constructions made in papers
\cite{ghosh}, \cite{cordero}. We hope to address this issue in more
detail in a future.

In the presence of the self-adjoint extension, the su(1,1) can no longer be implemented as the spectrum generating 
algebra as the dilatation generator in this case does not in general leave the domain of the Hamiltonian 
invariant \cite{jackiw, esteve, dh, we}. As a result, the spectrum for
a generic choice of the self-adjoint
 extension parameter 
is no  longer expressed in the Coulomb-like form. However, when $z = z_1$ or $z_2$, the
Coulomb-like nature of the spectrum is
 recovered and su(1,1) can again 
be implemented as a spectrum generating algebra.  This effect is
analogous to
 the quantum anomaly also observed in the pure 
Calogero type systems \cite{we, camblong}.
 
We have also seen that the system exhibits qualitatively different behaviour on two sides of the  point $ \alpha = 0.$
We find that for the attractive Coulomb potential
 ($ \alpha > 0$) there exists an infinite number of bound states. In
 the case of the repulsive Coulomb potential ($ \alpha < 0$), there
 appears to be at most a single bound state, which exists only for certain values of the
 self-adjoint extension parameter. 

In this paper we have restricted our discussion to the case when the coupling constant $g$ of the inverse square 
interaction is such that there is no collapse to the centre. It would be interesting to analyze this problem 
where the coupling is more attractive with $g < - \frac{1}{2}$, which would require renormalization group 
techniques \cite{jackiw, rajeev}.

\vskip 1cm

\noindent
{\bf Acknowledgment}\\

KSG would like to thank the School of Theoretical Physics at the
 Dublin Institute
 for Advanced Studies for hospitality, where a part of this work was done.
This work was supported by the Ministry of Science and Technology of
 the Republic
 of Croatia under contract No. 098-0000000-2865. 
This work was done within the framework of
the Indo-Croatian Joint Programme of Cooperation in Science and Technology
sponsored by the Department of Science and Technology, India (DST/INT/CROATIA/P-4/05), and
the Ministry of Science, Education and Sports, Republic of Croatia.



\begin{thebibliography}{99}

\bibitem{Calo}
  F.~Calogero,
  J.\ Math.\ Phys.\  {\bf 10}, 2191(1969);
  F.~Calogero,
  J.\ Math.\ Phys.\  {\bf 10}, 2197(1969);
  F.~Calogero,
  J.\ Math.\ Phys.\  {\bf 12}, 419(1971).

\bibitem{BS}
B. ~Sutherland, 
J.\ Math.\ Phys.\  {\bf 12}, 246(1971);
B. ~Sutherland, 
J.\ Math.\ Phys.\  {\bf 12}, 251(1971).

\bibitem{pr} M. A. Olshanetsky and A. M. Perelomov, Phys. Rep. {\bf 71}, 314
(1981); {\it ibid} {\bf 94}, 6 (1983).

\bibitem{poly} M. V. N. Murthy and R. Shankar, Phys. Rev. Lett. {\bf 73},
3331 (1994); Z. N. C. Ha, {\it Quantum Many-Body Systems in One
Dimension}, Series on Advances in Statistical Mechanics, Vol. 12, (World
Scientific, 1996);  A. P. Polychronakos, hep-th/9902157;
B. Basu-Mallick and A. Kundu, Phys. Rev. {\bf B62}, 9927
(2000).

\bibitem{qhe} H. Azuma and S. Iso, Phys. Lett. {\bf B331}, 107(1994).

\bibitem{ll} N. Kawakami and S.-K. Yang, Phys. Rev. Lett. {\bf 67}, 2493
(1991).

\bibitem{rmt} B. D. Simons, P. A. Lee, and B. L. Altshuler,
Phys. Rev. Lett. {\bf 72}, 64(1994); S. Jain, Mod. Phys. Lett. {\bf A11},
1201(1996).

\bibitem{qet} C. W. J. Beenakker and B. Rejaei, Phys. Rev. {\bf B49}, 7499
(1994); M. Caselle, Phys. Rev. Lett. {\bf 74}, 2776 (1995).

\bibitem{hs1} F. D. M. Haldane, Phys. Rev. Lett. {\bf 60}, 635 (1988);
B. S. Shastry, Phys. Rev. Lett. {\bf 60}, 639 (1988); A. P. Polychronakos,
Phys. Rev. Lett. {\bf 70}, 2329 (1993).

\bibitem{hs2} A. P. Polychronakos, Nucl. Phys. {\bf B419}, 553 (1994);  
B. Basu-Mallick, H. Ujino and M. Wadati, J. Phys. Soc. 
Jpn. {\bf 68}, 3219 (1999); F. Finkel, A. Gonzalez-Lopez, Phys. Rev. 
{\bf B72}, (2005) 174411; B. Basu-Mallick and N. Bondyopadhaya, Nucl. Phys.
{\bf B757}, 280 (2006).

\bibitem{sw} E. D'Hoker and D. H. Phong, hep-th/9912271; A. Gorsky and
A. Mironov, hep-th/0011197; A. J. Bordner, E. Corrigan and R. Sasaki, Prog.
Theor. Phys. {\bf 100}, 1107 (1998).

\bibitem{black}
G. W. Gibbons and P. K. Townsend, Phys. Lett. {\bf B454},
187 (1999).

\bibitem{brink}  A. P. Polychronakos, Phys. Rev. Lett. {\bf 69}, 703 (1992);
L. Brink, T. H. Hansson and M. A. Vasiliev, Phys. Lett.
{\bf B286}, 109 (1992); N. Gurappa and P. K. Panigrahi, Phys. Rev. {\bf B
59}, R2490 (1999).

\bibitem{Meljanac:2003jj}
  S.~Meljanac, M.~Milekovic and A.~Samsarov,
  Phys.\ Lett.\  B {\bf 573} 202 (2003);
  S.~Meljanac, M.~Milekovic and A.~Samsarov,
  Phys.\ Lett.\  B {\bf 594} 241 (2004);
  S.~Meljanac and A.~Samsarov,
  Phys.\ Lett.\  B {\bf 613} 221 (2005)
  [Erratum-ibid.\  B {\bf 620} 221 (2005)].


\bibitem{Meljanac:2006uf}
  S.~Meljanac, A.~Samsarov, B.~Basu-Mallick and K.~S.~Gupta,
  Eur.\ Phys.\ J.\  C {\bf 49} (2007) 875.

\bibitem{npb} 
   B. Basu-Mallick, Pijush K. Ghosh and Kumar S. Gupta, 
Nucl. Phys. {\bf B659}, 437 (2003).

\bibitem{we} B. Basu-Mallick and K. S. Gupta, Phys. Lett. {\bf A292},
36(2001);
B. Basu-Mallick, Pijush K. Ghosh and Kumar S. Gupta, 
Phys. Lett. {\bf A311}, 87 (2003).

\bibitem{feher} L. Feher, I. Tsutsui and T. Fulop,  
Nucl. Phys. {\bf B715}, 713 (2005); 
N. Yonezawa, I. Tsutsui, J. Math. Phys. {\bf 47}, 012104 (2006).

\bibitem{reed} M. Reed and B. Simon, {\it Methods of Modern Mathematical
Physics}, volume 2, (Academic Press, New York, 1972).

\bibitem{gerbert} P. Gerbert, Phys. Rev. {\bf D 40}, 1346 (1989).

\bibitem{jackiw} R. Jackiw in M.A.B. Beg Memorial Volume, A. ALi and P.
Hoodbhoy, eds. (World Scientific, Singapore, 1991).
\bibitem{manuel} C. Manuel and R. Tarrach, Phys. Lett. {\bf B 268}, 222
(1991); M. Bourdeau and R. D. Sorkin, Phys. Rev. {\bf D 45}, 687 (1992).

\bibitem{esteve} J. G. Esteve, Phys. Rev. {\bf D 34}, 674 (1986); J. G.
Esteve, Phys. Rev. {\bf D 66}, 125013 (2002).

\bibitem{falo} H. Falomir, P. A. G. Pisani and A. Wipf,
Jour. Phys. {\bf A 35}, 5427 (2002).

\bibitem{bal} C. Aneziris, A. P. Balachandran and Diptiman Sen, Int. Jour.
Mod. Phys. {\bf 6}, 4721 (1991).

\bibitem{trg} T. R. Govindarajan, V. Suneeta and S. Vaidya,
Nucl. Phys. {\bf B583}, 291 (2000); D. Birmingham, Kumar S. Gupta and
Siddhartha Sen, Phys. Lett. {\bf B505}, 191 (2001);
Kumar S. Gupta and Siddhartha Sen, Phys. Lett. {\bf B 526}, 121 (2002);
Kumar S. Gupta, hep-th/0204137.

\bibitem{khare}
  A.~Khare,
  J.\ Phys.\ A {\bf 29}, L45 (1996). 

\bibitem{ghosh} P.K. Ghosh and A. Khare, J. Phys. {\bf A32},  2129 (1999).   

\bibitem{abr} {\it Handbook of Mathematical Functions},
M. Abromowitz and I. A. Stegun (Dover Publications, New York, 1974).




\bibitem{borzov}
V. V. Borzov  and E. V. Damaskinsky, math.QA/0101215.

\bibitem{Dadic:2002qn}
  I.~Dadic, L.~Jonke and S.~Meljanac,
  Phys.\ Rev.\ D {\bf 67}, 087701(2003).

\bibitem{mmp}
S.~Meljanac, M.~Milekovi\'c and S.~Pallua,
Phys.\ Lett.\ B {\bf 328}, 55(1994).

  \bibitem{cordero} 
    P. Cordero and G. C. Ghirardy, Fortschr.Phys. {\bf 20}, 105(1972);
    M. Bernar, Ann.Phys. {\bf 75}, 305(1973);
    J. Ganster and B. Kohler, J.Phys.A {\bf 19}, 1291(1986).

\bibitem{fub} V. de Alfaro, S. Fubini and G. Furlan, Nuovo Cim. {\bf 34A},
569 (1976).

\bibitem{dh} E. D'Hoker and L. Vinet, Comm. Math. Phys. {\bf 97}, 391
(1985).

\bibitem{camblong} H. E. Camblong, L. N. Epele, H. Fanchiotti, C. A. Garcia Canal , 
Phys. Rev. Lett. {\bf 87}, 220402 (2001); H. E. Camblong, C. R. Ordonez, 
Phys. Rev. {\bf D 68}, 125013 (2003).  

\bibitem{rajeev} K. S. Gupta and  S. G. Rajeev,  Phys. Rev. {\bf D 48}, 5940 (1993). 
  
\end{thebibliography}
\end{document}